# A revisit of the papers on the theory of relativity: Reconsideration of the hypothesis of ether-dragging

Masanori Sato

*Honda Electronics Co., Ltd.,*

*20 Oyamazuka, Oiwa-cho, Toyohashi, Aichi 441-3193, Japan*

**Abstract:** This paper revisits previous papers related to the theory of relativity. Afterwards, a reconsideration of the hypothesis of ether-dragging is discussed. The ether is compatible with the theory of relativity and historical experiments; this paper explains the Michelson-Morley experiment using the ether-dragging hypothesis without the orthodox interpretation that the speed c is a fixed constant in terms of any system of inertial coordinates.

1. Introduction

The theory of special relativity, proposed by Einstein [1] in 1905, was simple and intuitive. The principle is the invariance of the speed of light: light in a vacuum propagates with the speed c (a fixed constant) regardless of the motion of the light source. However, the orthodox interpretation of the theory of special relativity was derived later from the results of the Michelson-Morley experiment [2], which suggested that "the speed c is a fixed constant in terms of any system of inertial coordinates". This orthodox interpretation is rather difficult to illustrate in the theory of special relativity: the Michelson-Morley experiment in the gravitational field of the earth cannot be discussed in the inertial coordinate; that is, the orthodox interpretation cannot be applied to the Michelson-Morley experiment. This is the starting point of this paper.

The idea that light in a vacuum propagates with the speed c (a fixed constant), regardless of the state of motion of the light source was commonly accepted by the end of the 19[th] and early 20[th] centuries; this idea was represented in Maxwell's equations and the wave equation as,

$$\frac{\partial^2 E}{\partial x^2} - \frac{1}{c^2}\frac{\partial^2 E}{\partial t^2} = 0, \tag{1}$$

where E is the amplitude of the wave, and c is the phase velocity of the wave. In those days, Maxwell and other scientists considered equation (1) to be defined in stationary coordinates. They considered that the speed of light, c, is defined in stationary coordinates, which is in the stationary ether. First of all, I would like to make this point clear: the orthodox interpretation is correct on the condition of the uniform flow of the ether; this is in the limitation of the theory of special relativity. Equation (1) has a solution $E \propto \exp i(\omega t - kx)$, (ω: frequency, *k*: wave number), which has a constant phase velocity $c = \omega / k$. This representation indicates that in the ether in uniform flow, the phase velocity is always constant. For example, the Doppler shift is detected as not only the frequency ω but also the wave number *k* that satisfy the constant phase velocity, c: this is because the wave number *k* is always proportional to the frequency ω in the inertial coordinate.

**Figure 1** (a) illustrates the idea that light in a vacuum propagates with the speed c regardless of the state of motion of the light source. This also illustrates an idea from Einstein's 1905 paper [1].

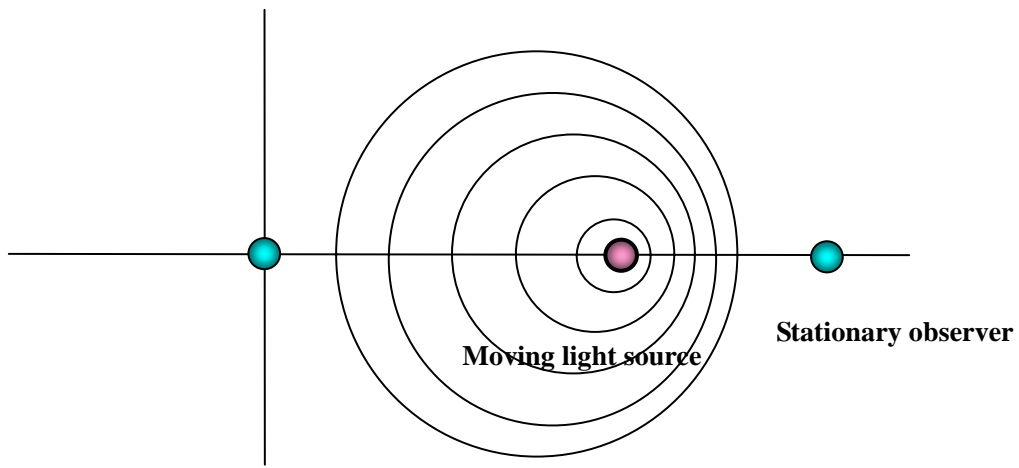

**Fig. 1** (a) Light in vacuum propagates with the speed c regardless of the state of motion of the light source

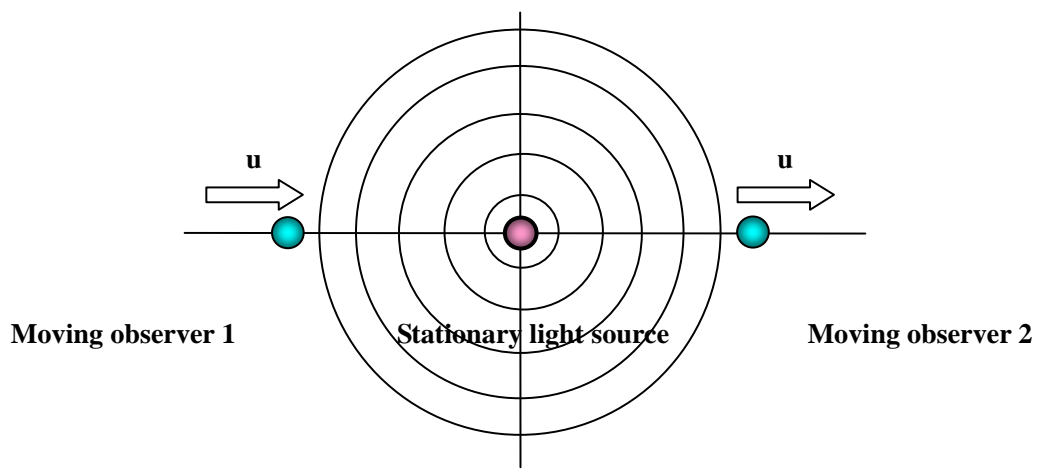

**Fig. 1** (b) Moving observers 1 and 2 detect the speed of light c

**Figure 1** (b) shows that moving observers 1 and 2 detect the constant phase velocity of light c regardless of their motion. The constant speed of light is satisfied on the condition that both the light source and the observer are in inertial motion. As far as wave equation (1) is discussed in the inertial coordinate system, the speed of light c is always constant. On the other hand, the interpretation of the theory of special relativity that the speed, c, is a fixed constant in terms of any system of inertial coordinates is rather ambiguous. This is due to the fact that when we discuss the ether-dragging, we have to assume the gravitational field. The ether-wind or ether-drift is not observed in the inertial coordinate system; the discussion should be carried out in the theory of general relativity.

To make the discussion more clear, let us consider an acoustic wave in the atmosphere, which has an isotropic constancy: although the earth moves in the solar system, we never consider the speed of the acoustic wave to be a fixed constant in any system of inertial coordinates. This is interpreted as the acoustic wave traveling in the atmosphere, which is completely dragged by the gravity of the earth. Although the motion of the earth in the solar system does not affect the speed of the acoustic wave, we never consider that the speed of an acoustic wave is a fixed constant in terms of any system of inertial coordinates. This is because, we know the gravitational field of the earth is not in the inertial coordinate system. Again, this paper starts from the simple question of how to illustrate the orthodox interpretation of this theory. I must conclude that this orthodox interpretation can be applied only in the case of a uniform ether. The frame of the wave equation does not appear to adhere to stationary coordinates; however, it does adhere to the physical constants of the medium. In the case of an acoustic wave in the atmosphere, the inertial coordinates of equation (1) are assumed to be the atmosphere. The physical coefficients are the density and the coefficient of stiffness. The wave equation of the electromagnetic wave can be interpreted using this analogy; that is, the inertial coordinates are the permittivity and the permeability around the earth.

As a counterargument for the orthodox interpretation, let us consider the earth-centered locally inertial (ECI) coordinate system in the global positioning system (GPS) experiment [3]. The reason why GPS works precisely in the ECI coordinate system is that the ECI coordinate system can be considered as the stationary state. It is difficult to calculate GPS in the solar system; only the ECI coordinate system works as the stationary frame. This experimental result is an analogy to an acoustic wave in the dragged-atmosphere that is, the electromagnetic wave in the dragged-ether by the gravity around the earth.

Using the analogy of an acoustic wave in the atmosphere, I use a classic hypothesis that the ether is the permittivity of free space, $\varepsilon_0$, and the permeability of free space, $\mu_0$. This classic hypothesis was derived from the proposal by Lorentz of luminiferous ether that the absolute stationary coordinate is defined in the stationary ether. Thus, to explain the Michelson-Morley experiment, he proposed the Lorentz contraction of length. I will also show that the complete ether-dragging hypothesis is compatible with the Michelson-Morley experiment. This hypothesis was derived from the proposal by Maxwell that the Maxwell equation and wave equation are satisfied in the stationary coordinate system, i.e., the stationary ether. Maxwell predicted an ether-wind; however, the GPS experiment showed that the ether-wind was not observed at least up to 20,000 km from the ground level. **Figure 2** shows that the ether is not only dragged, but also modified by gravity. The modification of the permittivity and the permeability by gravity causes a decrease in the speed of light, $c = \dfrac{1}{\sqrt{\varepsilon_0 \mu_0}}$. Diffraction around the gravitational potential of the sun, as observed by Eddington [4], can be explained using this proposal that light propagates toward regions of high refractive index, that is, toward the sun.

In the hypothesis that the ether is the permittivity and the permeability, the modification (increase) in the permittivity and permeability is used, rather than the curvature of spacetime.

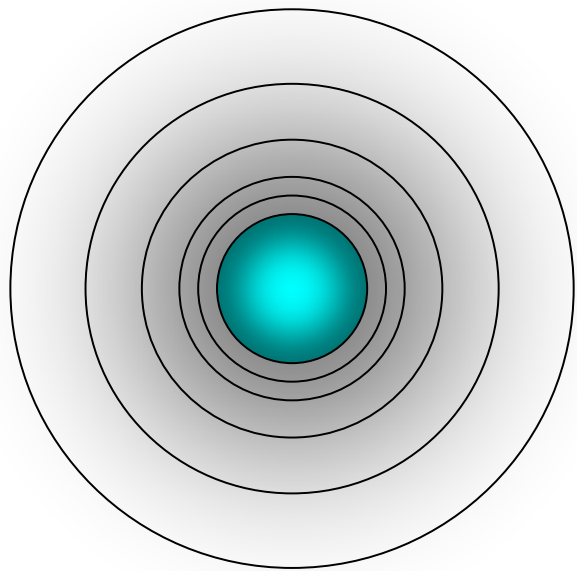

**Fig. 2** Gravitational field illustrated by the analogy of the atmosphere of the earth. The values of the permittivity of free space, $\varepsilon_0$, and permeability of free space, $\mu_0$, vary depending on the height. That is, the values are changed in order to satisfy the effect of the gravitational field on time dilation.

In the early 20th century, there were many great scientists who held very rigid beliefs in their own thoughts. Michelson accepted Einstein's work; however, he believed in the ether. He was not satisfied with the results of the Mt. Wilson experiment [2], and he repeated the experiment [5]. He worried that "shimmers" of air between the mountains might have fouled his results. In 1930,

Michelson's belief in the ether brought him to his last and most ambitious test [6], the measurement of the velocity of light in a partial vacuum. His daughter, Dorothy Michelson Livingston [7], wrote that Michelson never gave up his belief in the ether.

In 1924, Michelson-Gale-Pearson experiment [8] was carried out to observe the effect of the earth's rotation on the velocity of light. They assumed a fixed ether and the theory of special relativity. A fixed ether means the ether fixed to the ECI coordinate system; that is, the earth rotates in the ether. The theory of special relativity means that light in a vacuum propagates with the speed c regardless of the motion of the light source. They constructed the experimental setups using long pipes of partial vacuum. The experimental results showed the angular velocity of the earth in accordance with the theory of special relativity and the fixed ether. In those days, Michelson tried to prove the fixed ether experimentally; however the hypothesis of ether gradually disappeared.

In 1985, on the Sagnac experiment using GPS [9], there was no discussion of the ether. This is because Sagnac effect as well as Michelson-Gale-Pearson experiment can be reasonably explained without the hypothesis of ether.

Miller [10] was also a great scientist; he carried out the Michelson-Morley experiment with incredible enthusiasm. He was also a great experimentalist, and never changed his belief in the ether. In 1933, he reported experimental data that showed a slight seasonal and sidereal periodic fringe shift in the Michelson-Morley experiment. However, in 1955, his experimental results were re-evaluated and found to be thermal artifacts [11]. As far as the complete ether-dragging hypothesis is concerned, the null results are inevitable; thus, I believe that Miller's experimental results showed that the interferometer measurements are affected by the motion of the earth.

The null results of the GPS experiments were obtained by direct one-way measurement, which has a very high sensitivity compared to interferometer measurements. In a one-way (from the GPS satellite to the GPS station on earth) direct measurement of the speed of light, the sensitivity to a velocity of 30 km/s is calculated as $30 km/s \div 300,000 km/s = 1 \times 10^{-4}$. The sensitivity of the Michelson interferometer is estimated as

$$1 - \frac{1}{\sqrt{1 - \left(\frac{v_E}{c}\right)^2}} = 1 - \frac{1}{\sqrt{1 - \left(\frac{30}{300,000}\right)^2}} = 0.50 \times 10^{-8}.$$

Thus, the sensitivity of the direct one-way measurement is $2 \times 10^4$ higher than that of the Michelson interferometer. The null results are confirmed by one-way direct measurement in the GPS experiments.

At that time, it was hypothesized that ether-dragging occurs around the ground level. To check this hypothesis experimentally, the Michelson-Morley experiment was carried out using massive lead blocks (one path of the interferometer was set between two lead blocks); there was no fringe shift [12]. Michelson, Miller, and others discussed partial ether-dragging at the global magnetic field level; today, the GPS experiments show that if there is ether-dragging, it will be observed as an ether-wind more than 20,000 km from the ground level.

In 1951, Dirac [13, 14] referred to the ether in the context of his new electromagnetic theory. He suggested describing the ether from the viewpoint of quantum mechanics, that is, quantization of the ether. However, his interest in and discussion of the ether gradually disappeared. He wrote a book entitled "General Theory of Relativity" in 1975 [15], in which the ether was not described at all. As discussed by Dirac in the early 1950's, the ether may exhibit physical effects in quantum phenomena.

In this paper, a hypothesis of complete ether-dragging that is based on the beliefs of the great scientists is described. Thereafter, I will show that the historical experimental results are compatible with the ether. Although the hypothesis of no ether is compatible with the historical experimental results; however, they do not rule out the ether hypothesis.

2. Interpretation of historic experiments
2.1 Aberration of light is compatible with ether-dragging

The aberration of light was observed by Bradley in 1725. He explained the aberration using Newton's particle property of photons, as shown in **Fig. 3**. The aberration was considered to be one of the experimental results that show there is no ether-dragging around earth. Fresnel explained the aberration by assuming that the ether is unaffected by the motion of the earth [16]. This aberration is difficult to explain using the wave nature of the photon; however, it is easily explained using the particle nature of the photon.

Let us consider the hypothesis that the refractive index is dragged by the earth. The refractive index of air (n) is 1.000292, the speed of light (c) is 300,000 km/s, and the velocity of the earth in the solar system (v) is 30 km/s. Thus, using Fresnel's equation (2), we obtain

$$\frac{c}{n} + \left(1 - \frac{1}{n^2}\right)v \tag{2}$$

$$\therefore \frac{c}{n} + \left(1 - \frac{1}{n^2}\right)v = \frac{300,000}{1.000292} + \left(1 - \frac{1}{1.000292^2}\right)30 = 299,912 + 0.0175 \approx 299,912.$$

This calculation shows that the refractive index of the air modifies the speed of light. However, the contribution from the dragging of the refractive index by the earth is calculated as

$$0.0175 \div 299,912 = 0.5839 \times 10^{-9}.$$

For simplicity, let us assume a vacuum, that is, n=1. From Fresnel's equation (2), it can be concluded that the speed of light c is not affected by the velocity v. Therefore, we conclude that the aberration of light is clearly observed in the ether-dragging scenario.

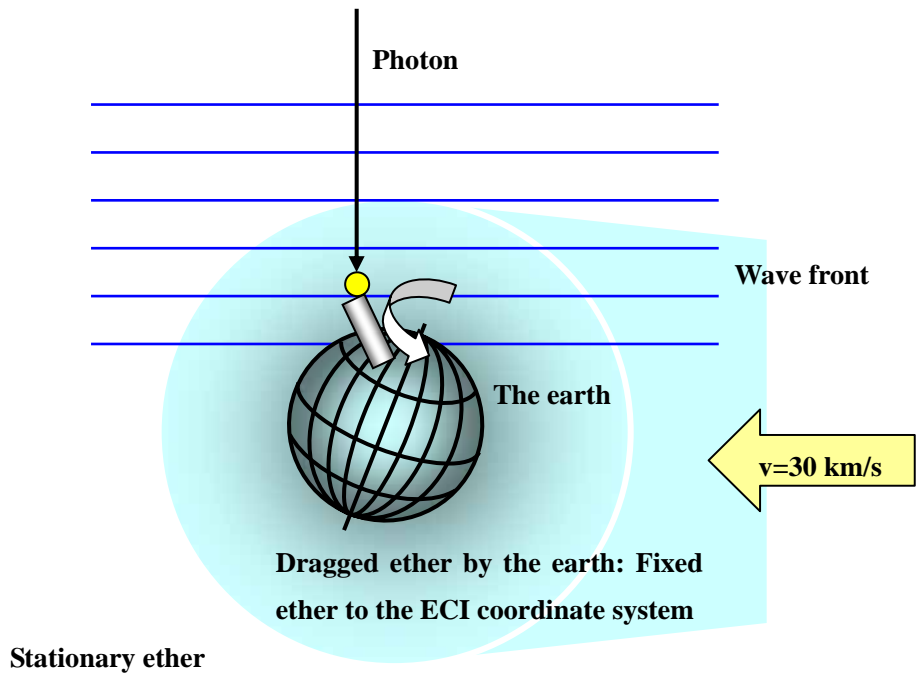

**Fig 3** Aberration of light is observed in the ether-dragging: Bradley explained the aberration using Newton's particle property of photons; however, it is difficult to explain using the wave nature of the photon. The ether is not only dragged but also fixed to the ECI coordinate system. The earth rotates in the fixed ether.

However, according to Fresnel's equation, the refractive index n depends on the frequency of the light. Of course, this dispersion is not observed, as the ether was considered to be frequency-independent. Thus, it was said that the aberration cannot be compatible with ether-dragging. This shows that it is difficult to interpret the aberration with the wave nature of the photon. The wave nature of light does not explain the compatibility between ether-dragging and the aberration. This is because the wave property shows that a photon is dragged by the ether.

Bradley explained the aberration using Newton's particle property of photons, which gives another simple illustration of a photon traveling in a straight line in the moving ether, without changing its direction, as shown in **Fig. 3**. Therefore, the aberration does not rule out ether-dragging. The particle property, in the particle-wave duality of the photon, makes the explanation simple.

2.2. Doppler shift

In 1842, Doppler proposed the Doppler shift of light in his treatise "*On the colored light of the binary stars and some other stars of the heaven*". If the star is moving toward us, the speed of a radiated photon is not changed by the motion of the light source; however, the photon has more energy and momentum, observed as the blue Doppler shift. It is interesting that the energy and momentum of a photon depend on the motion of the light source. It appears that the energy and momentum follow the Lorentz transformation, although the speed of the photon does not.

The energy and momentum of a photon depend on the motion of the light source: if the star is moving toward us, the photon has increased energy and momentum, which is the blue Doppler shift. If the observer is moving towards the light source, he detects a higher energy and momentum for the photon. Although the speed of a photon does not follow the Galilean transformation, the energy and momentum appear to follow the Lorentz transformation.

Let us discuss the Doppler shift of light. Equation (3) shows the longitudinal Doppler shift,

$$\nu_D = \nu_0 \sqrt{\frac{1 - \frac{u}{c}}{1 + \frac{u}{c}}}. \tag{3}$$

Here, $\nu_D$ is the Doppler frequency, and $\nu_0$ is the frequency of the source in the stationary state. To make the discussion simple, both the observer and the source are in free space; furthermore, either the observer or the source is in a stationary state, as shown in **Fig. 4**. The relative velocity is u. Thus, equation (3) can be used. The observer detects a Doppler frequency $\nu_D$.

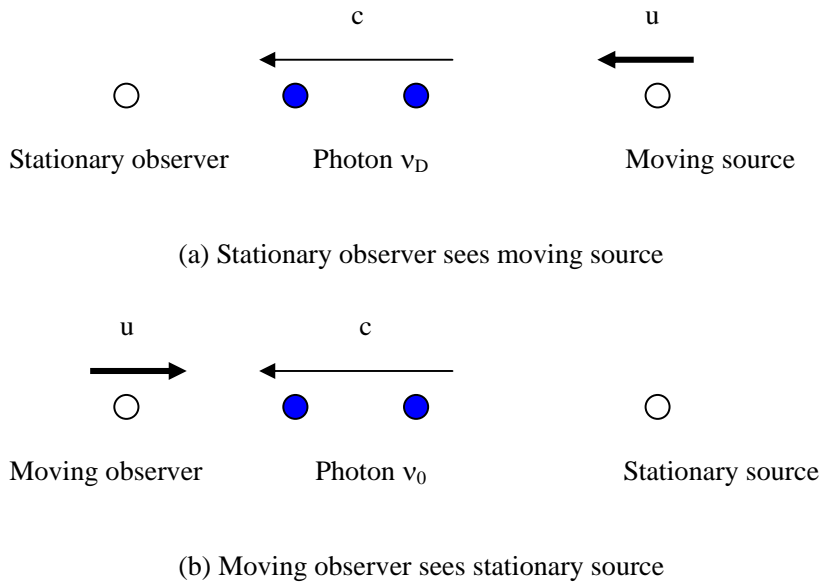

(a) Stationary observer sees moving source

(b) Moving observer sees stationary source

**Fig. 4** The Doppler shift from the viewpoint of quantum mechanics.
Both the observer and the source are in free space. Either the observer or the source is in a stationary state. The relative velocity is u. Thus, equation (3) can be used.

In **Fig. 4** (a), a moving source radiates a photon of energy $h\nu_D$ (h: Planck's constant); in **Fig. 4** (b), the moving observer detects a photon of energy $h\nu_D$. Although the speed of light, c, is constant, the moving source radiates a photon with higher energy and momentum. The moving observer then detects a photon with higher energy and momentum. In quantum mechanics, the phase velocity is defined as $c=\varepsilon/\mu$, ($\varepsilon$: energy, $\mu$: momentum), thus, the phase velocity is always constant.

Let us discuss the moving observer in **Fig. 4** (b) who detects a photon with speed c. It is possible to say that the moving observer detects a modified frequency and wave number as the Doppler shift. Thus, the phase velocity becomes a constant, that is, the speed of light, c. This interpretation is compatible with constant light speed, regardless of the motion of the light source and the orthodox interpretation.

The discussion of the Doppler shift assumes stationary coordinates. That is, when we use equation (3), it is assumed that either the observer or the source is in a stationary state.

2.3 Sagnac effect in GPS

Ashby [3] summarized the Doppler shift of carrier and modulated waves in GPS. He described that the Doppler shift of modulated wave is proportional to that of carrier. In the GPS, pulse coded modulation is used for the measurement of the distance. The Doppler shift of modulated wave is observed as the frequency change of modulated wave (i.e., wave packet). The Doppler shift of modulated wave is equivalent to the Sagnac effect in GPS. As shown in **Fig. 5**, the station on earth detects the Sagnac effect as well as the Doppler shift of modulated wave (Appendix).

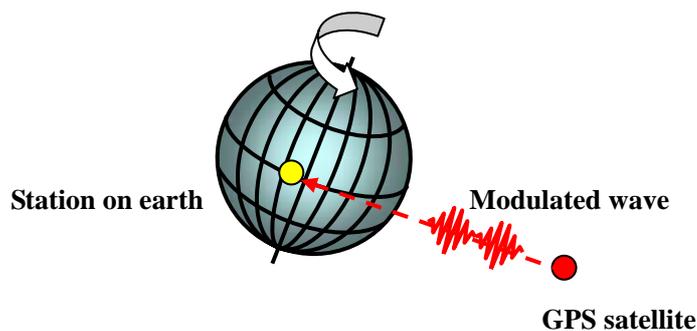

**Fig. 5** Sagnac effect and Doppler shift of modulated wave in GPS

Sagnac effect shows the distance change between the light source and the observer by the motion of the observer. If the observer moves on the flight time of light, the distance between the light source and the observer is changed. For example, let the distance between the GPS satellite (signal source) and the observer on earth set 30,000 km (the distance that light travels at 0.1 second). On the equator, the speed of the ground is around 0.47 km/s. Thus the Sagnac effect is 0.047 km at the measurement of 30,000 km for the observer on the equator, which means that at the flight time of light of 0.1 second, the observer moves 0.047 km. Sagnac experiment using the GPS showed the accuracy within 2 % [3, 9].

From the Sagnac effect, it is assumed that the ether has two properties: 1) the ether is dragged with the earth, 2) the ether does not rotate with the earth; the ether is fixed to the ECI coordinate system. Thus the earth rotates in the fixed ether as shown in **Fig. 5**. It looks like a comet. From these discussions, there remains two possible selections; one is there is no ether, and the other is fixed ether to the ECI coordinate system. The historical experiments do not rule out the ether hypothesis.

2.4 Summary of the historic experiments revisited

In this section, the previous experimental results are reconsidered with the hypothesis that the ether is the permittivity, $\varepsilon_0$, and the permeability, $\mu_0$ of free space (i.e., the refractive index n). Table 1 shows the historic experiments and their interpretation under the ether hypothesis.

Table 1 Experiments from historical papers

|   | Experiments | Comments |
|---|---|---|
| 1 | Aberration of light (1725) | The particle property of the photon gives a simple illustration; the photon travels in a straight line in the moving ether without changing direction. The aberration is compatible with the assumption of ether-dragging by the earth. |
| 2 | Doppler shift (1842) | The energy and momentum of the photon appear to satisfy the Lorentz transformation, although the speed of the photon does not. The moving light source causes a Doppler shift. The moving observer also detects the Doppler shift. The speed of light appears to be defined in the stationary frame; that is, the ether. |
| 3 | Michelson-Morley experiment [2] (1887) | In the 1887 experiment, Michelson and Morley ruled out a stationary ether. The experimental results (i.e., null results) can be interpreted as evidence for complete ether-dragging by the gravitational field of the earth. The Michelson-Morley experimental results can be explained using ether-dragging. |
| 4 | Sagnac (1913) | Sagnac effect is discussed in GPS-Sagnac effect. |
| 5 | Eddington [4] (1920) | Diffraction around the gravitational potential of the sun, as observed by Eddington, can be explained with the hypothesis that light propagates toward regions of high refractive index, that is, toward the sun. |
| 6 | Michelson-Gale -Pearson experiment [16] (1924) | Michelson-Gale-Pearson experiment was carried out to observe the effect of the earth's rotation on the velocity of light. They assumed a fixed ether and the theory of special relativity. The experimental results showed the angular velocity of the earth in accordance with the theory of special relativity and the fixed ether. |

| 7 | GPS-Sagnac [15] (1985) | Sagnac effect shows the distance change between the light source and the observer by the motion of the observer on the flight time of photon. The ether is dragged with the earth as well as fixed to the ECI coordinate system. The earth rotates in the ether. In the GPS, only the relative velocity defined in the ECI coordinate system is correct, we cannot use the relative velocity defined in the solar system. This experimental evidence reasonably explains the hypothesis of the fixed ether to the earth. |
|---|---|---|
| 8. | GPS [3] | GPS satellites orbit 20,000 km above ground level, at a velocity of 4 km/s. GPS uses an earth-centered locally inertial (ECI) coordinate system. This coordinate system can be assumed to be a stationary frame. In the GPS experiment, 20,000 km from ground level, the ether-wind and Fresnel's ether-dragging are not observed. The Michelson-Morley experiment conducted with GPS, i.e., by direct one-way measurement rather than with the use of a Michelson interferometer, confirms the constancy of the speed of light. The GPS experiments can be considered to be a reconfirmation of complete frame-dragging by the gravitational field of the earth. |

As described above, the complete ether-dragging hypothesis is compatible with the historical experimental results.

3. Ether-dragging and the stationary state

**Figure 6** illustrates ether-dragging by the gravitational fields of the earth and the sun. The ECI coordinate system (the gravitational field of the earth) drags the ether in the solar system. The solar system simultaneously drags the ether in the galaxy, and the galaxy also drags the ether in the cosmic microwave background (CMB). Thus, there are many stationary states: the ECI coordinate system, the galaxy and the CMB.

If we leave the gravitational field of the earth, we are in the gravitational field of the sun (solar system), that is, the sun-centered locally inertial coordinate system. When we travel at the speed of the earth (i.e., the relative velocity to the earth is 0) in the solar system, we observe a time dilation at a velocity of 30 km/s. If we reach the gravitational field of mars, we will be in another stationary state; that is, we will be in the mars-centered locally inertial coordinate system. When we leave the solar system, we will be in the galaxy; if we travel parallel to the solar system, we observe a time dilation caused by a speed of 230 km/s (the speed of the solar system in the galaxy). There are many stationary states in the solar system, for example, on the earth and mars. This is similar to the example of an acoustic wave, in which we have many stationary states, for example, in a moving train or in a flying airplane.

It is possible that ether-dragging occurs on larger scales (that is, the gravitational field scale) than were predicted by Michelson and Miller. They believed in the ether, and it is possible that their belief could be confirmed in space physics.

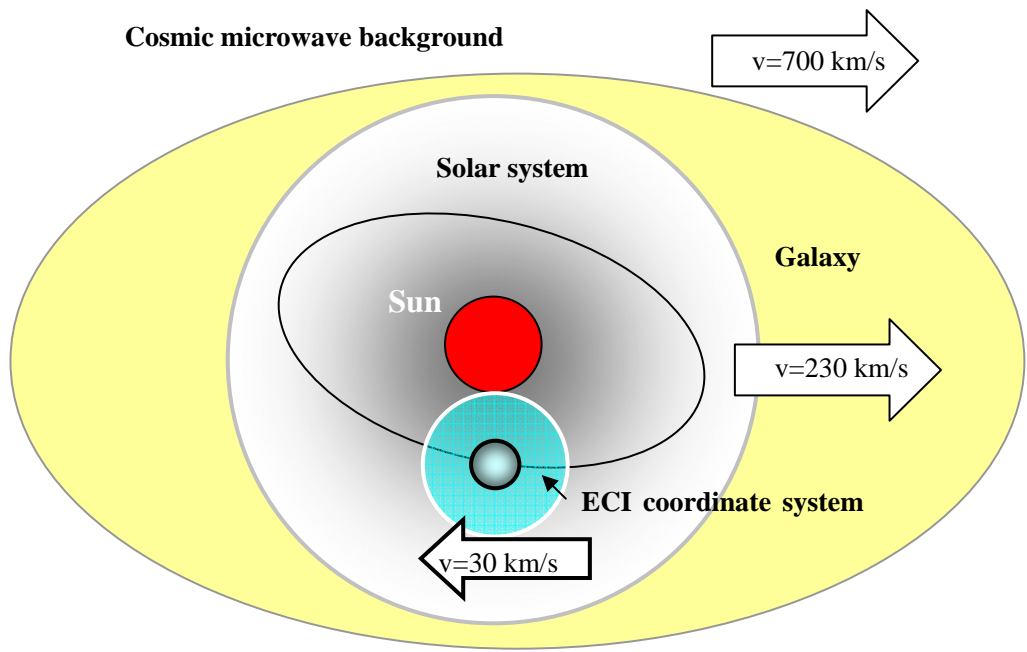

**Fig. 6** Illustration of the ether-dragging by the gravitational fields of the earth and the sun. Not only the earth but the sun also drags the ether in the galaxy, which moves in the cosmic microwave background (CMB). Thus, there are many stationary states; this is because the gravitational fields are the stationary states. For examples, the ECI coordinate system, the solar system, the galaxy and the CMB.

4. Discussion

4.1 The wave-particle duality in the theory of relativity

In the above discussions, I used the wave and particle properties. It is not easy to explain the experimental results using only one property. For example, the aberration is explained using the particle property, and I cannot explain the aberration with the wave property. On the other hand, the Doppler shift is difficult to explain only by the particle property; the wave property is needed. In the interpretation of the theory of relativity, particle-wave duality is required.

Thus, the ether-dragging hypothesis is explained using the wave-particle duality; I cannot consistently explain this hypothesis using only one of the two properties. Therefore, the ether-dragging hypothesis should be checked experimentally with space physics.

4.2 Merits and drawbacks of the ether

Let us discuss the merits and faults of the ether hypothesis. The faults of the ether hypothesis are that it is unnecessary as far as explaining observed phenomena, and that it has never been observed; however, it is harmless and eases the understanding of the theory of special relativity. The concept of the ether is very convenient; it defines a stationary reference frame. Thus, a relative velocity is also defined.

As described above, one of the candidates for the ether is the permittivity of free space, $\varepsilon_0$, and the permeability of free space, $\mu_0$. These are very familiar from electromagnetic theory; however, the physical meanings of the permittivity and the permeability have not been studied thoroughly enough.

The merit of the ether hypothesis is that a stationary reference frame can be defined. This stationary reference frame is compatible with the theory of special relativity; however, it is not compatible with the interpretation of special relativity. I only disagree with the interpretation, which is very strong and usually used in the discussion of the gravitational field of the earth.

At first, I considered the ether to be an additional concept to aid in understanding the theory of special relativity; at this stage, however, I have come to believe in the physical reality of the ether. I believe that the ether is compatible with the theory of relativity; the Michelson-Morley experiment can be solved using the theory of general relativity without the orthodox interpretation.

## 5. Conclusion

The consideration of the ether began because I could not illustrate the expression "the speed, c, is a fixed constant in any system of inertial coordinates." In this study, I found that many great scientists believed in the ether. It was surprising for me to read their splendid papers. The ether is compatible with the theory of relativity; furthermore, I think the ether is useful for understanding the theory of relativity. Although the ether is said to be superfluous, it appears to have a physical reality. I believe that searching for the ether opens up possibilities for new physics. For example, the quantization of the ether, as pointed out by Dirac, is very attractive. There is a possibility that experiments in space physics may obtain new results regarding the ether. Today, we have new information from quantum mechanics and the GPS experiments. In the days of these highly valuable historic papers, the authors did not have such information. Therefore, it is important to revisit historic papers from the viewpoint of modern physics.

**Appendix**: Sagnac effect on the group velocity

Let us discuss the group velocity change depending on the observer's velocity. Sagnac effect is observed as the distance change between the light source and the observer by the motion of the observer. Let us consider two observers on the equator. The two observers 1 and 2 are combined by a rigid rod of length L as shown in **Fig. A**. Not only observer 1 but also observer 2 observes the Sagnac effect. That is, observer 2 moves during the flight time of light between observers 1 and 2. Thus, observer 2 also detects Sagnac effect of $1.57 \times 10^{-6}$; which indicates that a light reaches earlier to the observers than in the stationary state.

Considering the Lorentz transformation, the group velocity is calculated as,

$$c_G = \frac{L \times 1/\gamma}{\Delta t_S \times 1/\gamma} = \frac{L}{\Delta t_S} = \frac{L}{t_2 - t_1}. \tag{A-1}$$

$$\gamma = \frac{1}{\sqrt{1-\left(\frac{v}{c}\right)^2}}.$$

Where, $\gamma$ is the Lorentz factor, $t_1$ is the time when a light reaches observer 1, $t_2$ is the time when a light reaches observer 2, and $\Delta t_S = t_2 - t_1$. Let the differential time of stationary observers to set $\Delta t_0$, thus we obtain,

$$c = \frac{L}{\Delta t_0}.$$

The Lorentz factor appears both in length and time; therefore it is cancelled as shown in equation (2).

**Figure B** shows the Sagnac effect between observers 1 and 2; after observer 1 detects the coded wave, observer 2 moves towards the GPS satellite. At the time $t_2$, observer 2 moves $\Delta t_S \times v$. From **Fig. B** we obtain,

$$\Delta t_S \times c = L - \Delta t_S \times v. \tag{A-2}$$

$$\therefore \Delta t_S = \frac{L}{c+v}. \tag{A-3}$$

$$c_G = \frac{L}{\Delta t_S} = \frac{L}{\Delta t_0}\left(1 + \frac{v}{c}\right) = c + v. \tag{A-5}$$

The Sagnac effect in GPS data shows group velocity change due to the motion of the observer.

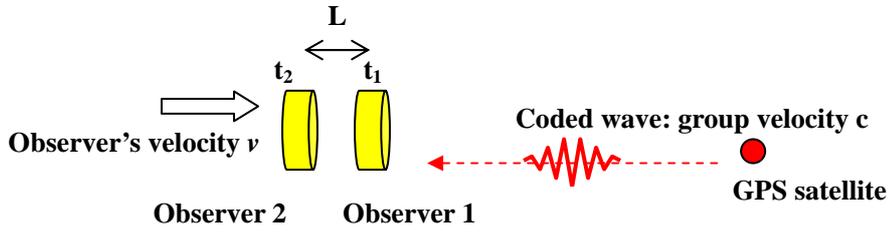

**Fig. A** Sagnac effect on the group velocity measurement using GPS: pulse coded signal is detected by observers 1 and 2. The detected times $t_1$ and $t_2$ suffer the Sagnac effect. Thus, the differential time $\Delta t_S = t_2 - t_1$ becomes smaller than $\Delta t_0$ in which observers 1 and 2 are in the stationary states.

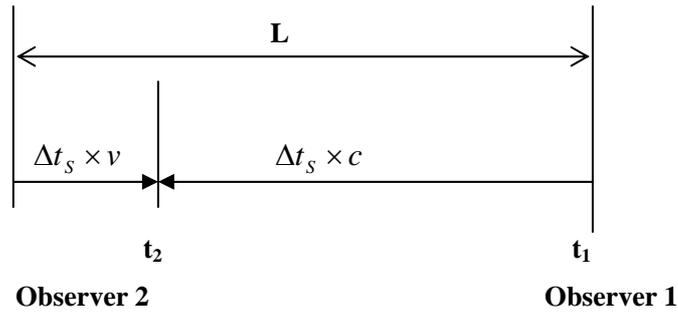

**Fig. B** Derivation of the equation of Sagnac effect: after observer 1 detects the coded wave, observer 2 moves towards the GPS satellite. At the time $t_2$, when the coded wave reaches at observer 2, which moves $\Delta t_S \times v$. Thus, we obtain $\Delta t_S \times c = L - \Delta t_S \times v$.